\documentclass[aps, prl, twocolumn,superscriptaddress, groupedaddress]{revtex4-2}
\usepackage[utf8]{inputenc}
\usepackage{amsmath}
\usepackage{amsthm}
\usepackage{amsfonts}
\usepackage{times,txfonts}
\usepackage{braket}
\usepackage{graphicx}
\usepackage{soul}
\usepackage[english]{babel}
\usepackage{color}
\usepackage[hypertexnames, breaklinks=true, bookmarksnumbered=true, bookmarksopen=true, colorlinks=true, linktocpage=true, citecolor=magenta, urlcolor=magenta, linkcolor=magenta]{hyperref}%
\newtheorem{theorem}{Theorem}
\definecolor{cadetblue}{rgb}{0.37, 0.62, 0.63}
\definecolor{orange}{rgb}{1,0.54,0}
\DeclareMathOperator{\tr}{Tr}

\begin{document}

\title{Entanglement and coherence in Bernstein-Vazirani algorithm}
\author{Moein Naseri}
\author{Tulja Varun Kondra}
\author{Suchetana Goswami}
\author{Marco Fellous-Asiani}
\author{Alexander Streltsov}
\affiliation{Centre for Quantum Optical Technologies, Centre of New Technologies, University of Warsaw, Banacha 2c, 02-097 Warsaw, Poland}

\begin{abstract}
Quantum algorithms allow to outperform their classical counterparts in various tasks, most prominent example being Shor's algorithm for efficient prime factorization on a quantum computer. It is clear that one of the reasons for the speedup is the superposition principle of quantum mechanics, which allows a quantum processor to be in a superposition of different states at the same time. While such superposition can lead to entanglement across different qubits of the processors, there also exists quantum algorithms which outperform classical ones using superpositions of individual qubits without entangling them. As an example, the Bernstein-Vazirani algorithm allows one to determine a bit string encoded into an oracle. While the classical version of the algorithm requires multiple calls of the oracle to learn the bit string, a single query of the oracle is enough in the quantum case. In this Letter, we analyze in detail the quantum resources in the Bernstein-Vazirani algorithm. For this, we introduce and study its probabilistic version, where the goal is to guess the bit string after a single call of the oracle. We show that in the absence of entanglement, the performance of the algorithm is directly related to the amount of quantum coherence in the initial state. We further demonstrate that a large amount of entanglement in the initial state prevents the algorithm from achieving optimal performance. We also apply our methods to quantum computation with mixed states, proving that pseudopure states achieve optimal performance for a given purity in the Bernstein-Vazirani algorithm. We further investigate quantum resources in the one clean qubit model, showing that the model can exhibit speedup over any known classical algorithm even with arbitrary little amount of multipartite entanglement, general quantum correlations, and coherence.

\end{abstract}

\maketitle

\textit{Introduction.} The non-local nature of the quantum correlations \cite{EPR_35, S_35, B_64, WJD_07, JWD_07,HorodeckiRevModPhys.81.865} is one of the key features which enables quantum states outperform the classical ones in different information processing tasks such as quantum teleportation and different cryptographic protocols~\cite{BBCJPW_93, BW_92, BB_84, E_91}. In present days, it is quite evident that these non-classical correlations play the role of the resources in quantum computers which are more powerful than classical computers in performing certain algorithms. For example, the Deutsch-Jozsa algorithm \cite{DJ_92} and Shor's algorithm \cite{S_94} show exponential speedup over the best known classical algorithm \footnote{For Deutsch-Jozsa algorithm, there is however no speedup if one uses a probabilistic classical computer \cite{johansson2017efficient}}, when implemented on a quantum computer. Similarly quantum search algorithms, such as Grover's algorithm \cite{G_96}, offer a quadratic speed up over the classical ones \cite{montanaro2016quantum}. Other than these, there are many quantum algorithms based on query complexity, showing potential speed up over the classical computers~\cite{A_07, LMRSS_11}.

Even though entanglement plays a crucial role in many quantum computational tasks~\cite{LP_01, JL_03, V_13}, it is yet to be concluded whether entanglement is in general necessary to obtain some kind of quantum advantage, especially when noise is taken into account. Indeed, various results suggest that in certain setups quantum computers can outperform their classical counterparts also without entanglement, for some specific definition of quantum advantage. For instance, it has been shown that the Deutsch-Jozsa and Simon algorithms, when implemented on mixed states which remain separable at all times, induce non-classical features for the output state~\cite{BIHAM200415}. Calling the oracle in the quantum regime only once provides a small -- but non-zero -- amount of information about the computational task. On the other hand, a single call of the oracle in the classical case provides absolutely no information. Hence, despite the fact it does not show that an exponential advantage is kept in presence of separable states, it indicates that quantum computation without entanglement can be more powerful than classical computing. The one clean qubit model~\cite{KL_98} allows for efficient estimation of the normalized trace of an $n$-qubit unitary which can be implemented efficiently in terms of quantum gates, showing exponential speedup over the best known classical algorithm~\cite{DFC_05}. The algorithm operates on highly mixed quantum states, exhibiting a bounded amount of entanglement across any bipartition~\cite{DFC_05}. The question of whether or not quantum entanglement is necessary at all in a general case in order to obtain a quantum speedup in the one clean qubit model is still an open question \cite{DattaPhysRevLett.100.050502,lanyon2008experimental,datta2007role,Dakic2010,passante2012measuring}. In~\cite{GFE_09}, it has been shown that states with a large value of geometric entanglement are not often useful for speedup in measurement-based quantum computation. Moreover, it has been proven that universal quantum computation is possible even with arbitrary little entanglement~\cite{V_13}.

In this Letter, we investigate quantum resources in the Bernstein-Vazirani (BV) algorithm, which allows to identify an unknown bit string $\boldsymbol a$ encoded as a linear function in an oracle \cite{BV_97,BGK_18}. While it is not possible to obtain complete information about the bit string by calling the oracle only once in the classical case, in the quantum domain a single call of the oracle is enough for this purpose~\cite{BV_97}. We introduce the \emph{probabilistic Bernstein-Vazirani algorithm}, where the goal is to guess the bit string $\boldsymbol a$ after a single call of the oracle with maximal probability. While the BV algorithm does not require entanglement in principle \cite{BV_97}, our methods allow for a rigorous quantitative investigation of entanglement and coherence in the protocol. We estimate the maximal guessing probability for all pure initial states, and show that without entanglement in the initial and the final states the performance is directly related to the amount of coherence in the initial state.  

For the BV algorithm operating on mixed states, we investigate the role of purity for the performance. We give a closed expression for the maximal probability to guess the bit string as a function of purity, and also provide the quantum states achieving optimal performance. Our results reveal that optimal performance for a given amount of purity is achieved for pseudopure states, which are useful for NMR quantum computing~\cite{BCJLPS_99, LP_01}. These results suggest that NMR is a suitable platform for implementing the BV algorithm, supporting earlier experiments in this direction~\cite{DSZFYHW_01}. Another type of quantum algorithms which is relevant for NMR quantum computing is the one clean qubit model~\cite{KL_98}. For this model, we show that a large class of quantum resource and correlation quantifiers can be made arbitrarily small, without influencing the performance of the algorithm for normalized trace estimation. This includes widely used quantifiers of multipartite entanglement, general quantum correlations, quantum coherence, and mutual information.

\medskip
\emph{Coherence and entanglement quantification in multipartite systems}. We will now present quantifiers of coherence and entanglement which will be used in this Letter. Given an incoherent reference basis $\{\ket{i}\}$, the amount of coherence of a state $\rho$ can be quantified via the robustness of coherence \cite{NBCPJA_16,PCBNJA_16}
\begin{equation}
\label{robustness_coherence}
R(\rho)=\min_{\tau}\left\{ s\geq0:\frac{\rho+s\tau}{1+s} \in \mathcal I \right\}.
\end{equation}
Here, $\tau$ is a density matrix and $\mathcal I$ is the set of incoherent states, i.e., quantum states which are diagonal in the reference basis. A similar measure can also be defined for quantum entanglement~\cite{VidalPhysRevA.59.141,SteinerPhysRevA.67.054305} and general quantum resource theories, where it has an operational interpretation via channel discrimination tasks~\cite{RegulaPhysRevLett.122.140402,RegulaPhysRevX.9.031053}. For bipartite systems, the reference basis is naturally defined as $\{\ket{i}\ket{j}\}$, where $\{\ket{i}\}$ and $\{\ket{j}\}$ are the incoherent bases of the individual subsystems. Extension to multipartite systems is done in a similar fashion~\cite{BromleyPhysRevLett.114.210401,StreltsovPhysRevLett.115.020403,SAP_17}.

To quantify entanglement in multipartite systems, we will use distance-based measures of the form~\cite{VPRK_97,VedralPhysRevA.57.1619,HorodeckiRevModPhys.81.865}
\begin{equation}
E(\rho)=\inf_{\sigma \in \mathcal{S}}D(\rho,\sigma), \label{eq:EntanglementMeasure}
\end{equation}
where $\mathcal S$ is the set of separable states. $D$ is a (pseudo-)distance fulfilling $D(\rho,\sigma)\geq 0$ with equality if and only if $\rho = \sigma$, and the data processing inequality: $D(\Lambda[\rho],\Lambda[\sigma]) \leq D(\rho,\sigma)$ for any quantum operation $\Lambda$. An important example for a distance with such properties is the quantum relative entropy $S(\rho||\sigma) = \tr[\rho \log_2 \rho] - \tr[\rho \log_2 \sigma]$, and the corresponding entanglement measure is known as the relative entropy of entanglement~\cite{VPRK_97,VedralPhysRevA.57.1619}. Another entanglement measure which will be used in this Letter is the geometric entanglement~\cite{Shimony1995,WG_03,Streltsov_2010}, which is obtained from Eq.~(\ref{eq:EntanglementMeasure}) by choosing $D(\rho,\sigma) = 1 - F(\rho,\sigma)$ with fidelity $F(\rho,\sigma) = \{\tr[(\sqrt \rho \sigma \sqrt \rho )^{1/2}]\}^2$. We note that quantum coherence can also be quantified with measures of the form~(\ref{eq:EntanglementMeasure}), if the set of separable states $\mathcal S$ is replaced by the set of incoherent states $\mathcal I$~\cite{BaumgratzPhysRevLett.113.140401,SAP_17}.

\medskip
\emph{Probabilistic Bernstein-Vazirani algorithm.}
The goal of the BV algorithm \cite{BV_97} is to find an unknown $N$-bit string $\boldsymbol{a}=a_{1}, \ldots, a_{N}$ with $a_i \in \{0,1\}$ encoded as a linear function 
\begin{equation}
    f(\boldsymbol{x})=\boldsymbol{a} \cdot \boldsymbol{x} \!\!\!\! \mod 2= \left(\sum_{k=0}^{N}a_{k}x_{k}\right) \!\!\!\!\! \mod 2
\end{equation}
on the $N$-bit string $\boldsymbol{x}=x_{1}, \ldots, x_{N}$. In particular, one aims to find the string $\boldsymbol a$ with a minimal number of queries of the function $f$. Classically, the optimal strategy is to evaluate $f$ for each input $\boldsymbol x$ where one of the bits is set to 1, and all the other $N-1$ bits are set to 0, resulting in $N$ queries of the function $f$ \cite{BV_97}. 

However in quantum domain, we only need to make one call of the function to learn the bit string $\boldsymbol a$ \cite{BV_97}. For this, we make the standard assumption that the bit string $\boldsymbol{x}$ is encoded into a $N$-qubit quantum state of the form $\ket{\boldsymbol{x}}=\otimes_{i=1}^N\ket{x_i}$. Moreover, we assume that the function is encoded into an oracle, which is represented by a unitary $U_{\boldsymbol a}$ acting on states of the form $\ket{i}\ket{\boldsymbol{x}}$ with $i\in \{0,1\}$ as follows:
\begin{equation}
    U_{\boldsymbol a}(\ket{i}\ket{\boldsymbol{x}}) = \ket{i \oplus f(\boldsymbol x)} \ket{\boldsymbol x},
\end{equation}
and $\oplus$ denotes addition modulo 2. In the following, the first qubit will be called \emph{oracle register}, whereas the remaining $N$ qubits will be called \emph{system qubits}. If the oracle unitary $U_{\boldsymbol a}$ is applied onto the state $\ket{-} \ket{+}^{\otimes N}$  with $\ket{\pm} = (\ket{0} \pm \ket{1})/\sqrt 2$, the final state is given by $\sum_{\boldsymbol{x}}(-1)^{\boldsymbol a \cdot \boldsymbol{x}}\ket{-}\ket{\boldsymbol{x}}/2^N$. Discarding the oracle register and applying a Hadamard gate on each of the qubits, the overall state is converted into $\ket{\boldsymbol{a}}$. Finally, the bit string $\boldsymbol a$ can be obtained by measuring each of the qubits in the computational basis~\cite{BV_97}.

So far we have seen that the BV algorithm shows optimal performance if the initial state is $\ket{-} \ket{+}^{\otimes N}$. We will now analyze the performance of the algorithm for general input states. In general, we cannot expect that the procedure works optimally if the initial state is different from $\ket{-} \ket{+}^{\otimes N}$. To capture the performance in the general setup, we assume that we have no prior knowledge about the bit string $\boldsymbol a$, i.e., each of the possible bit strings can occur with equal probability. Since there are $2^N$ possible bit strings, the probability of each bit string is given by $1/2^N$. In order to learn $\boldsymbol a$, we allow to apply the oracle unitary $U_{\boldsymbol a}$ onto a general quantum state $\rho$, and perform a general quantum measurement onto the final state $U_{\boldsymbol a}\rho U_{\boldsymbol a}^\dagger$. We call this protocol \emph{probabilistic Bernstein-Vazirani algorithm}.

The performance of the probabilistic BV algorithm can now be defined via the optimal probability to guess the bit string $\boldsymbol a$, corresponding to the maximal probability to guess the oracle unitary $U_{\boldsymbol a}$. This is analogous to the definition for the average guessing probability in channel discrimination tasks, see e.g.~\cite{RegulaPhysRevX.9.031053}. For a set of quantum channels $\Lambda_i$, each applied with probability $p_i$, channel discrimination can be achieved by applying the channel onto an initial quantum state $\rho$, and performing a positive operator-valued measure (POVM) $\{M_i\}$ on the final state. The average probability for correctly guessing the channel is then given by $\sum_{i}p_{i}\tr[\Lambda_{i}(\rho)M_{i}]$. We are now ready to define the performance of the probabilistic BV algorithm as the maximal probability to guess the bit string $\boldsymbol a$:
\begin{equation}
P(\rho)=\frac{1}{2^N}\max_{\{M_{\boldsymbol a}\}}\sum_{\boldsymbol a}\tr\left[U_{\boldsymbol a} \rho U_{\boldsymbol a}^{\dagger}M_{\boldsymbol a}\right].
\label{eq:Psucc_oracle}
\end{equation}

Having defined the performance of the BV algorithm for general input states, we will now provide a closed expression for the performance for all pure initial states. For this, note that every pure state of $N+1$ qubits can be written as 
\begin{equation}
    \ket{\mu} = a \ket{+}\ket{\phi'} + b \ket{-}\ket{\phi}, \label{eq:GeneralInitialState}
\end{equation}
where $\ket{\phi}$ and $\ket{\phi'}$ are states of $N$ qubits. In the following, $c_{\boldsymbol {x}}$ denote the coefficients of the state $\ket{\phi}$ in the computational basis, i.e., $\ket{\phi} = \sum_{\boldsymbol{x}} c_{\boldsymbol{x}} \ket{\boldsymbol{x}}$.
\begin{theorem}
The performance of the probabilistic BV algorithm for a pure initial state is given as
\begin{eqnarray}
   P (\ket{\mu}) &&= \frac{1}{2^{N}}
  \Bigg[ 1+|b|^2R(\ket{\phi}\!\bra{\phi}) \nonumber\\
   &&+2 |b|\sum_{x\neq {\boldsymbol{0}}}|c_{\boldsymbol{x}}|\bigg(\sqrt{1-|b|^2\bigg(1-|c_{{\boldsymbol{0}}}^2|\bigg)}-|b||c_{{\boldsymbol{0}}}|\bigg)\Bigg].
   \label{BV-OracleReg}
\end{eqnarray}
where $R$ is the robustness of coherence in the computational basis.
\label{theorem_1}
\end{theorem}
 We refer to the Supplemental Material for the proof of the theorem. As we further discuss in the Supplemental Material, maximal performance $P(\ket{\mu})=1$ is achievable only if $\ket{\mu} = \ket{-}\ket{\psi_{\max}}$, where $\ket{\psi_{\max}}$ is a maximally coherent state of $N$ qubits. Moreover, we can generalize the result of the theorem to a class of mixed quantum states. We assume that we initiate the BV algorithm in the mixed state of the form $\rho=\sum_{i}p_{i}\ket{\mu_{i}}\!\bra{\mu_{i}}$ with $\ket{\mu_{i}}=a_{i}\ket{+}\ket{\phi}+b_{i}\ket{-}\ket{\psi_{i}}$, $|a_{i}|^2+|b_{i}|^2=1$, and $\langle00...0|\psi_{i}\rangle=0$. For this class of mixed states the performance of the algorithm is given by
\begin{equation}
    P(\rho)=\frac{1+R'(\rho)}{2^{N}},
\end{equation}
where $R'$ is the robustness of coherence in the basis $\{\ket{+}\ket{\phi}\}\cup \{\ket{-}\ket{\boldsymbol{x}}\}$. We refer to the Supplemental Material for more details.

We will now compare the probabilistic version of the BV algorithm presented above to its classical counterpart. For this, we first define the classical version of the probabilistic BV algorithm. In the classical case, the BV algorithm transforms the $N+1$ bit string $(i,\boldsymbol x)$ into $(i \oplus f(\boldsymbol x),\boldsymbol x)$. Assuming that each of the possible functions $f$ is applied with equal probability $1/2^N$, the performance of the classical BV algorithm can be defined as the probability for guessing the bit string $\boldsymbol a$ if the algorithm is applied onto the bit string $(i,\boldsymbol x)$. As we will see in the Supplemental Material, the probability for correctly guessing $\boldsymbol a$ is given by
\begin{equation} \label{Classical-Performance}
P_\mathrm{c}(\boldsymbol{x})=\begin{cases}
\frac{1}{2^{N}} & \mathrm{if}\,\boldsymbol{x}=\boldsymbol{0},\\
\frac{1}{2^{N-1}} & \mathrm{otherwise}.
\end{cases}
\end{equation}
Comparison of Eqs.~(\ref{Classical-Performance}) and (\ref{BV-OracleReg}) shows that in the classical version of the algorithm it is not possible to achieve the performance higher than $1/2^{N-1}$ in one call of the oracle. A better performance is possible in the quantum case, where optimal performance $P=1$ is achievable for some initial states.

\medskip
\textit{Probabilistic Bernstein-Vazirani algorithm without entanglement.} As can be seen from Theorem \ref{theorem_1}, quantum coherence plays an important role in the performance of the BV algorithm. The performance of the algorithm is given explicitly in Eq. (\ref{BV-OracleReg}) and here the total initial state $\ket{\mu}$ can be an entangled one or a separable one. As entanglement is considered an expensive resource in quantum information theory~\cite{HorodeckiRevModPhys.81.865} it is reasonable to investigate the performance of the algorithm in the absence of entanglement between all the $N+1$ qubits both before and after the action of the oracle. 

We will first focus on pure initial states, and extend our discussion to mixed states below. As we show in the Supplemental Material, for the probabilistic BV algorithm to exhibit nontrivial performance above $1/2^N$~\footnote{Note that performance $1/2^N$ can be achieved for any initial state.} without entanglement in the initial and the final state, the total initial state must have the form
\begin{equation}
\ket{\mu}=\ket{-}\ket{\phi} \label{eq:NoEntanglement}
\end{equation}
with an $N$-qubit product state $\ket{\phi}$.  The action of the unitary $U_{\boldsymbol a}$ on such states can be written as 
\begin{equation}
U_{\boldsymbol{a}}\left(\ket{-}\ket{\phi}\right)=\ket{-}\otimes\left(V_{\boldsymbol{a}}\ket{\phi}\right)
\end{equation}
where the $N$-qubit unitary $V_{\boldsymbol a}$ can be implemented by applying $\sigma_z$ on $i$-th qubit conditioned on the value of $a_i$, i.e., $V_{\boldsymbol a}=\otimes_{i=1}^{N}\sigma_{z,i}^{a_{i}}$. Note that $V_{\boldsymbol a}$ does not create entanglement in the $N$-qubit system. 

To be certain that entanglement does not play any role in the algorithm, we will also check whether the optimal POVM $\{M_{\boldsymbol a}\}$ in Eq.~\eqref{eq:Psucc_oracle} is implementable without using entanglement. While the density matrix right before applying the POVM is a mixture of non-entangled states that we need to distinguish to deduce the bit-string $\boldsymbol a$, it could still be that performing the POVM maximizing Eq.~\eqref{eq:Psucc_oracle} requires non-local operations~\cite{bennett1999quantum,PhysRevLett.122.040403}. In the Supplemental Material we show that the optimal POVM is indeed implementable without entanglement. 

As we directly see from Theorem~\ref{theorem_1}, for states of the form~(\ref{eq:NoEntanglement}) the performance can be given as $P(\ket{-}\ket{\phi}) = [1 + R(\ket{\phi})]/2^N$. This result applies regardless whether the $N$-qubit state $\ket{\phi}$ is product or not, and also extends to mixed states:
\begin{equation}
    P\left(\ket{-}\!\bra{-}\otimes\sigma\right)=\frac{1+R(\sigma)}{2^N}, \label{eq:BVQubits}
\end{equation}
we refer to the Supplemental Material for more details. The relation Eq. (\ref{eq:BVQubits}) suggests an operational meaning for the robustness of coherence of $\rho$ in terms of quantum computation. We note that in general the POVM achieving optimal performance for mixed states as in Eq.~(\ref{eq:BVQubits}) requires entanglement to be implemented.

\medskip
\emph{Multipartite entanglement in Bernstein-Vazirani algorithm.} As we have seen so far, entanglement is not required to reach optimal performance. If the initial state is given by $\ket{-}\ket{+}^{\otimes N}$, it is possible to perfectly learn the bit string $\boldsymbol a$ with a single application of the oracle unitary. We will now go one step further, and investigate how multipartite entanglement between the $N$ system qubits influences the performance of the probabilistic BV algorithm. We investigate the relation between robustness of coherence and geometric entanglement for $N$-qubit systems and show that a large amount of geometric entanglement in the system state can be disadvantageous for the performance.

For this, we focus on $N$-qubit W-states~\cite{dur2000three}:
\begin{align}
    \ket{\Psi_W} = \frac{1}{\sqrt{N}}(e^{i\phi_{1}}\ket{\psi_1}\ket{\psi_2} \ldots \ket{\psi_N^\perp}+e^{i\phi_{2}}\ket{\psi_1} \ldots \ket{\psi_{N-1}^\perp}\ket{\psi_N}\nonumber\\
    + \ldots +e^{i\phi_{N}}\ket{\psi_1^\perp}\ket{\psi_2} \ldots \ket{\psi_N}),
\end{align}
where $\ket{\psi_{i}}$ and $\ket{\psi_{i}^\perp}$ are orthogonal. Note that for $N=3$, the W-states are the only type of states achieving maximal geometric entanglement~\cite{TamaryanPhysRevA.80.052315}. As we show in the Supplemental Material, a W-state for $N>2$ is never maximally coherent. For $N=3$ this implies that for initial states of the form $\ket{-}\ket{\psi}$ there is a threshold on the geometric entanglement of $\ket{\psi}$ above which it is not possible to reach optimal performance $P=1$. This result is similar in spirit to the results presented in~\cite{GFE_09}, showing that quantum states can be too entangled to be useful for quantum computation. 

\medskip
\emph{Purity in probabilistic Bernstein-Vazirani algorithm.} We will now apply the results presented above, obtaining the optimal performance of the BV algorithm in the presence of noise. For this, we assume that the initial state of the algorithm has bounded purity, i.e., $\tr[\rho^2] \leq \gamma$. The following theorem provides the optimal initial states in this case.
\begin{theorem} \label{thm:Pseudopure}
Having the oracle register in the state $\ket{-}$, the optimal initial state of the $N$-qubit system maximizing the performance of the BV algorithm with bounded purity $\tr[\rho^2] \leq \gamma$ is given by
\begin{equation}
   \rho_{\max,\gamma}=\frac{d}{2\lambda_{1}}\ket{\psi_{\max}}\!\bra{\psi_{\max}}+\frac{\lambda_{2}}{2\lambda_{1}}\openone
\end{equation}
with 
\begin{align}
\lambda_{1} & =\frac{d\sqrt{1-\frac{1}{d}}}{2\sqrt{\gamma-\frac{1}{d}}},\,\,\,\,\,\,\,\,\lambda_{2}=\frac{\sqrt{1-\frac{1}{d}}}{\sqrt{\gamma-\frac{1}{d}}}-1,
\end{align}
and $\ket{\psi_{\max}}$ being a maximally coherent state. 
The optimal performance in this case is given as
\begin{equation}
P(\rho_{\max,\gamma})=\frac{1}{d}+\frac{d-1}{2\lambda_{1}}
\end{equation}
where $d=2^N$.
\end{theorem}

\noindent We refer to the Supplemental Material for the proof.

As we see, the state maximizing the robustness of coherence for a bounded fixed amount of purity is a pseudopure state which is possible to prepare with NMR technique~\cite{LP_01, cory1997ensemble,sharf2000spatially}. Also, as the performance $P(\rho)$ is monotonically related to the robustness of coherence, we conclude that NMR quantum computing is a suitable platform for implementation of the probabilistic BV algorithm.

Several results presented in this Letter also apply for a generalized version of the BV algorithm where, instead of qubits, we have particles with a $D$-dimensional Hilbert space. More details can be found in the Supplemental Material.
\medskip

\emph{Quantum resources and the power of one qubit.} Earlier we have seen that NMR is a suitable platform for implementing the BV algorithm. Another quantum computational model that is often discussed in the context of NMR is deterministic quantum computation with one qubit (DQC1)~\cite{KL_98}. Here, the initial state is given by $\rho\otimes \openone_N/2^N$, with a single-qubit state $\rho = (\openone + \alpha \sigma_x)/2$, maximally mixed $N$-qubit state $\openone_N/2^N$, and parameter $0 \leq \alpha \leq 1$. As discussed in~\cite{DFC_05}, it is possible to efficiently estimate the normalized trace of an $N$-qubit unitary $U_N$ by applying a controlled version of it to the initial state, with the first qubit as the control and the remaining $N$ qubits as the target:
\begin{equation}
    V_N = \ket{0}\!\bra{0} \otimes \openone + \ket{1}\!\bra{1} \otimes U_N.
\end{equation}
We assume that $V_N$ can be implemented efficiently in terms of quantum gates. By performing a measurement on the first qubit of the final state, it is possible to efficiently estimate the normalized trace $\tr(U_N)/2^N$ whenever $\alpha > 0$~\cite{DFC_05}. As of today, no efficient classical algorithm for solving this problem is known~\cite{DFC_05}.

Various works have tried to identify the reason for the quantum speedup in this task by analyzing properties of the quantum states of the processor in the algorithm~\cite{DFC_05,DattaPhysRevLett.100.050502,Dakic2010,Matera2015}. The amount of bipartite entanglement in the final state has been analyzed in~\cite{DFC_05}. The authors investigate entanglement generated by the DQC1 algorithm in different bipartitions, as quantified by negativity~\cite{ZyczkowskiPhysRevA.58.883,VidalPhysRevA.65.032314}. The authors of~\cite{DFC_05} conclude that negativity is bounded by a constant, which is independent of the number of qubits. Motivated by these findings, it has been suggested that other types of quantum correlations, such as quantum discord~\cite{ModiRevModPhys.84.1655,Streltsov2014}, are responsible for the speedup, since a typical instance of the algorithm exhibits nonzero quantum discord in a certain bipartition~\cite{DattaPhysRevLett.100.050502}. Nevertheless, there is evidence for exponential speedup even without discord~\cite{Dakic2010}, and it has been argued that the performance of trace estimation with DQC1 is rather related to quantum coherence in the algorithm~\cite{Matera2015}.

We will now show that normalized trace estimation with DQC1 can be implemented efficiently even though a very general class of quantum resource and correlations quantifiers is arbitrarily small in every step of the algorithm. We will start our analysis with general quantifiers of entanglement, and extend it to other measures below. We will show that any measure of entanglement of the form~(\ref{eq:EntanglementMeasure}) is bounded by a constant in the DQC1 protocol. For this, note that the maximally mixed state of $N+1$ qubits $\openone_{N+1}/2^{N+1}$ is fully separable, and thus $E(\sigma) \leq D(\sigma,\openone_{N+1}/2^{N+1})$ for any $N+1$-qubit state $\sigma$. Recall that the initial state of the DQC1 protocol is given by $\rho \otimes \openone_N/2^N$ with a qubit state $\rho$. After the application of $V_N$ we obtain
\begin{align}
E\left(V_{N}\rho\otimes\frac{\openone_{N}}{2^{N}}V_{N}^{\dagger}\right) & \leq D\left(V_{N}\rho\otimes\frac{\openone_{N}}{2^{N}}V_{N}^{\dagger},\frac{\openone_{N+1}}{2^{N+1}}\right) \label{eq:EntanglementDQC1}\\
 & =D\left(\rho\otimes\frac{\openone_{N}}{2^{N}},\frac{\openone_{N+1}}{2^{N+1}}\right)=D\left(\rho,\frac{\openone_1}{2}\right). \nonumber
\end{align}
Here, we used the fact that any distance $D$ which fulfills the data processing inequality is invariant under unitaries and under attaching or discarding ancillary systems, i.e., $D(\rho,\sigma) = D(U \rho U^{\dagger},U\sigma U^{\dagger})$ and
$D(\rho,\sigma) = D(\rho \otimes \tau,\sigma \otimes \tau)$.

Recalling that DQC1 allows for efficient estimation of the normalized trace of $U_N$ for any $\rho = (\openone + \alpha \sigma_x)/2$ as long as $\alpha > 0$~\cite{DFC_05}, we see that multipartite entanglement in the algorithm is bounded by a constant as long as $D(\rho,\openone/2) \leq c$ for all qubit states $\rho$. For any continuous distance $D$ this constant can be chosen arbitrarily small by choosing $\alpha$ appropriately. 

Note that the form of the unitary $V_N$ is not relevant in Eq.~(\ref{eq:EntanglementDQC1}), the result holds for any unitary acting on the total $N+1$ qubit state. Because of this, it applies also to the intermediate states of the algorithm $\rho_i$, which are the states of the quantum processor after the application of $i$ quantum gates. This proves that normalized trace estimation with DQC1 is possible with an arbitrary small amount of multipartite entanglement at all times.

The method presented above does not only apply to entanglement, but to a general class of quantum resource and correlation quantifiers which vanish on maximally mixed states. To see this, let us consider a general quantity of the form
\begin{equation}
    \mathcal{M}(\rho)=\inf_{\sigma \in \mathcal{F}} D(\rho, \sigma), \label{eq:DQC1genralmeasures}
\end{equation}
where $\mathcal F$ is some set of $N+1$-qubit states containing the maximally mixed state, and $D$ is a distance with the properties discussed above. It is immediately clear that the arguments from Eq.~(\ref{eq:EntanglementDQC1}) apply to any such quantity. To see that the above results apply to the quantum mutual information 
\begin{equation}
I\left(\rho^{AB}\right)=S\left(\rho^{A}\right)+S\left(\rho^{B}\right)-S\left(\rho^{AB}\right),
\end{equation}
recall that the mutual information can also be written as in Eq.~(\ref{eq:DQC1genralmeasures}), if $D$ is the quantum relative entropy and $\mathcal F$ is the set of product states \cite{BertaPhysRevLett.121.190503}. Here, the systems $A$ and $B$ can be any subsets of the $N+1$ qubits. The results also apply to various measures of quantum correlations beyond entanglement~\cite{ModiRevModPhys.84.1655,Streltsov2014}, if $\mathcal F$ is the set of classically correlated states $\rho_{\mathrm{cc}}=\sum_{i,j}p_{ij}\ket{a_{i}}\!\bra{a_{i}}\otimes\ket{b_{j}}\!\bra{b_{j}}$  or classical-quantum states $\rho_{\mathrm{cq}}=\sum_{i}p_{i}\ket{a_{i}}\!\bra{a_{i}}\otimes\sigma_i$ with local orthonormal bases $\{\ket{a_i}\}$ and $\{\ket{b_j}\}$ and general local states $\sigma_i$. Our results also apply to the relative entropy of coherence~\cite{BaumgratzPhysRevLett.113.140401}, if $\mathcal F$ is the set of incoherent states and $D$ is the quantum relative entropy. Finally, if $\mathcal F$ contains only the maximally mixed state of $N+1$ qubits, the quantifier $\mathcal M$ becomes a measure of purity~\cite{HorodeckiPhysRevA.67.062104,GOUR20151,Streltsov2018}. 

In summary, we see that efficient normalized trace estimation with DQC1 is possible even if the protocol exhibits an arbitrary small amount of multipartite entanglement, mutual information, general quantum correlations, coherence, or purity in every step of the computation.

\medskip
\emph{Conclusion.} In this work we have introduced and studied the probabilistic version of the Bernstein-Vazirani algorithm, where the goal is to optimally guess a bit string $\boldsymbol a$ encoded into an oracle unitary. We have evaluated the optimal performance of the algorithm for all pure initial states, using the maximal guessing probability of the bit string to quantify the performance. For the case that no entanglement is present in the initial and the final state of the algorithm, we show that the performance is directly related to the amount of coherence in the initial state. We also show that a large amount of geometric entanglement can prevent the algorithm from reaching optimal performance. 

Our methods are also applicable to quantum computation with mixed initial states. For the probabilistic Bernstein-Vazirani algorithm operating on noisy states, we show that pseudopure states achieve optimal performance for a given amount of purity. Since pseudopure states are useful in NMR quantum computation, our results suggest that NMR is a suitable platform for the probabilistic Bernstein-Vazirani algorithm. We further analyze quantum features in the DQC1 model, allowing to estimate the normalize trace of an $n$-qubit unitary which can be implemented efficiently in terms of quantum gates. We show that the DQC1 circuit can be implemented efficiently with a vanishingly small amount of quantum resources and correlations. This applies to a general class of resource and correlation quantifiers, including multipartite entanglement, mutual information, quantum coherence, and purity. These results suggest that the reason for the speedup of DQC1 is unlikely to be rooted in the properties of the quantum states of the processor.

\medskip
\emph{Acknowledgements.} This work was supported by the ``Quantum Optical Technologies'' project, carried out within the International Research Agendas programme of the Foundation for Polish Science co-financed by the European Union under the European Regional Development Fund and the ``Quantum Coherence and Entanglement for Quantum Technology'' project, carried out within the First Team programme of the Foundation for Polish Science co-financed by the European Union under the European Regional Development Fund.

\bibliography{BV.bib}

\section{Supplemental Material}

\subsection{Proof of Theorem \ref{theorem_1}}
To prove the result in Theorem \ref{theorem_1}, we first prove Eq.~(\ref{eq:BVQubits}) in the main text. The robustness of asymmetry of a given density matrix $\rho$ is defined as below \cite{PCBNJA_16}:
\begin{equation}
    R_A(\rho)=\min_{\tau}\left\{ s\geq0:\frac{\rho+s\tau}{1+s} \in \mathcal F\right\},
\end{equation}
in which $F$ is the set of symmetric states with respect to action of a group. As indicated in \cite{PCBNJA_16}, the robustness of asymmetry can be expressed as a semidefinite program (SDP). As the resource theory of quantum coherence can be seen as a special case of resource theory of asymmetry by considering the symmetry with respect to $U(1)$ group \cite{PCBNJA_16}, we can extrapolate the SDP for the robustness of coherence and it can be expressed as the following SDP maximization over $X$~\cite{PCBNJA_16, NBCPJA_16}:
\begin{subequations} \label{eq:SDP}
\begin{align}
   R(\rho)=\underset{X}{\textrm{max }}[\tr(\rho X)-1]\\
    X\geq 0\\
    E(X)=\openone
\end{align}
\end{subequations}
where $E(X)=\frac{1}{d}\sum_{\boldsymbol{a}}u_{\boldsymbol{a}}Xu_{\boldsymbol{a}}^{\dagger}$ and $u_{\boldsymbol{a}}=\sum_{\boldsymbol{x}}e^{i\frac{2\pi}{d}\boldsymbol{a}\boldsymbol{x}}\ket{\boldsymbol{x}}\!\bra{\boldsymbol{x}}$. 

In order to prove the expression in Eq.~(\ref{eq:BVQubits})
of the main text we will show that $P(\ket{-}\!\bra{-} \otimes \rho)$ is both lower and upper bounded by $[1+R(\rho)]/d$ with $d=2^N$. As discussed in the main text, the action of the oracle unitaries $U_{\boldsymbol a}$ onto states of the form $\ket{-}\!\bra{-} \otimes \rho$ can be expressed as 
\begin{equation}
U_{\boldsymbol{a}}\ket{-}\!\bra{-}\otimes\rho U_{\boldsymbol{a}}^\dagger=\ket{-}\!\bra{-}\otimes V_{\boldsymbol{a}}\rho V_{\boldsymbol{a}}^{\dagger}
\end{equation}
with the $N$-qubit unitaries $V_{\boldsymbol a}=\otimes_{i=1}^{N}\sigma_{z,i}^{a_{i}}$. The performance of the probabilistic BV algorithm can thus be written as 
\begin{equation}
P(\ket{-}\!\bra{-} \otimes \rho)=\max_{\{M_{\boldsymbol a}\}}\frac{1}{d}\sum_{\boldsymbol a}\tr\left(V_{\boldsymbol a} \rho V_{\boldsymbol a}^{\dagger}M_{\boldsymbol a}\right).
\end{equation}
To prove Eq.~(\ref{eq:BVQubits}) of the main text, we will now show that 
\begin{equation}
\max_{\{M_{\boldsymbol a}\}}\frac{1}{d}\sum_{\boldsymbol a}\tr\left(V_{\boldsymbol a} \rho V_{\boldsymbol a}^{\dagger}M_{\boldsymbol a}\right)\geq\frac{1+R(\rho)}{d}.
\end{equation}
For this, we define the following operators:
\begin{equation}
    M_{\boldsymbol a}^{\prime}=\frac{1}{d}V_{\boldsymbol a}XV_{\boldsymbol a}^{\dagger},
\end{equation}
where $X$ is the operator maximizing the SDP in Eqs.~(\ref{eq:SDP}). The operators $M_{\boldsymbol a}^{\prime}$ are positive because $X$ is positive. Since $E(X)=\openone$, we have 
\begin{equation}
\frac{1}{d}\sum_{k=1}^{d}X_{\boldsymbol{x},\boldsymbol{x}}=1,
\label{XDiagElement}
\end{equation}
where $X_{\boldsymbol{x},\boldsymbol{y}}$ are the components of the matrix $X$. Thus, the diagonal elements of $X$ are the same and equal to $1$. Noting that $\frac{1}{d}\sum_{\boldsymbol{a}}e^{i\frac{2\pi}{D} \boldsymbol{a}\cdot(\boldsymbol{x}-\boldsymbol{x'})}=\delta_{\boldsymbol{x},\boldsymbol{x'}}$ and $V_{\boldsymbol a}$ are diagonal unitaries, we have
\begin{align}
\sum_{\boldsymbol a}M_{\boldsymbol a}^{\prime}&=\frac{1}{d}\sum_{f}V_{\boldsymbol a}XV_{\boldsymbol a}^{\dagger} \\
&=\sum_{\boldsymbol{x},\boldsymbol{y}}\frac{1}{d}\sum_{\boldsymbol{a}}e^{i\frac{2\pi}{D}\boldsymbol{a}\cdot(\boldsymbol{x}-\boldsymbol{y})}X_{\boldsymbol{x},\boldsymbol{y}}\ket{\boldsymbol{x}}\!\bra{\boldsymbol{y}}=\openone. \nonumber
\end{align}
Therefore $\{M_{\boldsymbol a}'\}$ forms a POVM. Thus, we must have
\begin{align}
\frac{1}{d}\sum_{\boldsymbol{a}}\tr\left(V_{\boldsymbol{a}}\rho V_{\boldsymbol{a}}^{\dagger}M_{\boldsymbol{a}}^{\prime}\right) & =\frac{1}{d}\sum_{\boldsymbol{a}}\frac{1}{d}\tr\left(V_{\boldsymbol{a}}\rho V_{\boldsymbol{a}}^{\dagger}V_{\boldsymbol{a}}XV_{\boldsymbol{a}}^{\dagger}\right)\label{GEQ}\\
 & =\frac{1}{d}\tr(\rho X)=\frac{1+R(\rho)}{d}\nonumber \\
 & \leq\underset{\{M_{\boldsymbol{a}}\}}{\max}~\frac{1}{d}\sum_{\boldsymbol{a}}\tr\left(V_{\boldsymbol{a}}\rho V_{\boldsymbol{a}}^{\dagger}M_{\boldsymbol{a}}\right). \nonumber 
\end{align}
Above we use the definition of the  robustness of coherence to obtain $\tr(\rho X) = 1+R(\rho)$. 

Now, we are going to indicate that $\big(1+R(\rho)\big)/d$ is also an upper bound for the performance. From the definition of the robustness of coherence, it follows that 
\begin{equation}
    \rho=[1+R(\rho)]\sigma-R(\rho) \tau,
\end{equation}
with some quantum state $\tau$ and some incoherent state $\sigma$. Hence, for any POVM $\{M_{\boldsymbol a}\}$ we have,
\begin{align}
\sum_{\boldsymbol{a}}\tr\left(V_{\boldsymbol{a}}\rho V_{\boldsymbol{a}}^{\dagger}M_{\boldsymbol{a}}\right) & =[1+R(\rho)]\sum_{\boldsymbol{a}}\tr\left(V_{\boldsymbol{a}}\sigma V_{\boldsymbol{a}}^{\dagger}M_{\boldsymbol{a}}\right)\\
 & -R(\rho)\sum_{\boldsymbol{a}}\tr\left(V_{\boldsymbol{a}}\tau V_{\boldsymbol{a}}^{\dagger}M_{\boldsymbol{a}}\right)\nonumber \\
 & \leq[1+R(\rho)]\sum_{\boldsymbol{a}}\tr\left(V_{\boldsymbol{a}}\sigma V_{\boldsymbol{a}}^{\dagger}M_{\boldsymbol{a}}\right).\nonumber 
\end{align}
As $\sigma$ is an incoherent state, it holds $V_{\boldsymbol a}\sigma V_{\boldsymbol a}^{\dagger}=\sigma$ and
\begin{equation}
\sum_{\boldsymbol{a}}\tr\left(V_{\boldsymbol{a}}\sigma V_{\boldsymbol{a}}^{\dagger}M_{\boldsymbol{a}}\right)=\tr\left(\sigma\sum_{\boldsymbol{a}}M_{\boldsymbol{a}}\right)=1.
\end{equation}
Thus, we arrive at the inequality
\begin{equation}
\max_{\{M_{\boldsymbol{a}}\}}\frac{1}{d}\sum_{\boldsymbol{a}}\tr\left(V_{\boldsymbol{a}}\rho V_{\boldsymbol{a}}^{\dagger}M_{\boldsymbol{a}}\right)\leq\frac{1+R(\rho)}{d}.\label{LEQ}
\end{equation}
From Eqs.~\eqref{GEQ} and \eqref{LEQ} we have
\begin{equation} \label{performance-d}
\max_{\{M_{\boldsymbol a}\}}\frac{1}{d}\sum_{\boldsymbol a}\tr\left(V_{\boldsymbol a}\rho V_{\boldsymbol a}^\dagger M_{\boldsymbol a}\right)=\frac{1+R(\rho)}{d}.
\end{equation}
This completes the proof of Eq.~(\ref{eq:BVQubits}) of the main text.

Now we prove the result in Theorem~\ref{theorem_1}. Recall that the oracle acts as follows:
\begin{equation}
    \ket{\mu}=a \ket{+} \ket{\phi'} + b \ket{-} \ket{\phi} \rightarrow \ket{\mu_{\boldsymbol a}}=a \ket{+} \ket{\phi'} + b \ket{-} \ket{\psi_{\boldsymbol a}}.
    \label{oraclereg}
\end{equation}
We now reorder the final state $\ket{\mu_{\boldsymbol a}}$ in a suitable manner. Let us consider the computational basis $\{\ket{\boldsymbol{x}}\}_{\boldsymbol{x}}$. We have
\begin{equation}
    \ket{\psi_{\boldsymbol a}}=\sum_{\boldsymbol{x}} c_{\boldsymbol{x}} (-1)^{\boldsymbol{a}\cdot \boldsymbol{x}} \ket{\boldsymbol{x}},
\end{equation}
where $c_{\boldsymbol{x}}$ is the coefficient of the state $\ket{\phi} = \sum_{\boldsymbol x} c_{\boldsymbol x}\ket{\boldsymbol x}$. For convenience we denote $\boldsymbol{0}=00...0$ (the string with $N$ zeros).
Hence,
\begin{equation}
    \ket{\mu_{\boldsymbol a}}=a \ket{+} \ket{\phi'}+b \ket{-} (c_{\boldsymbol{0}} \ket{0}^{\otimes N}) +b \ket{-}  \left(\sum_{\boldsymbol{x}\neq \boldsymbol{0}} c_{\boldsymbol{x}} (-1)^{\boldsymbol{a}\cdot \boldsymbol{x}}\ket{\boldsymbol{x}}\right)
\end{equation}
Here the dimension of the whole system along with the oracle register is considered to be $2^{N+1}$. Now,
\begin{eqnarray}
    \ket{\mu_{\boldsymbol a}}= \sqrt{|a|^2+|b|^2 |c_{\boldsymbol{0}}|^2} \ket{\psi'_{\boldsymbol{0}}} + b   \sum_{\boldsymbol{x}\neq \boldsymbol{0}} c_{\boldsymbol{x}} (-1)^{\boldsymbol{a}\cdot \boldsymbol{x}}\ket{\psi_{\boldsymbol{x}}'})
\end{eqnarray}
with the states,
\begin{equation}
\ket{\psi'_{\boldsymbol{0}}}=\frac{a\ket{+}\ket{\phi'}+bc_{\boldsymbol{0}}\ket{-}\ket{0}^{\otimes N}}{\sqrt{|a|^{2}+|b|^{2}|c_{\boldsymbol{0}}|^{2}}}
\end{equation}
and $\ket{\psi'_{\boldsymbol{x}}}= \ket{-} \ket{\boldsymbol{x}}$ for $\boldsymbol x \neq \boldsymbol 0$. Note that the state $\ket{\psi_{\boldsymbol{x}}'}$ is orthogonal to the state $\ket{\psi_{\boldsymbol{0}}'}$ for all $\boldsymbol{x} \neq \boldsymbol 0$. Constructing a new basis with these orthogonal states, we can rewrite the final state as, 
\begin{equation}
    \ket{\mu_{\boldsymbol a}} = \sum_{\boldsymbol{x}} c'_{\boldsymbol{x}} (-1)^{\boldsymbol{a}\cdot \boldsymbol{x}} \ket{\psi'_{\boldsymbol{x}}}
    \label{mu_i}
\end{equation}
with $c'_{\boldsymbol{0}}= \sqrt{|a|^2+|b|^2 |c_{\boldsymbol{0}}|^2}$ and $c'_{\boldsymbol{x}} = b c_{\boldsymbol{x}}$ for $\boldsymbol{x}\neq \boldsymbol{0}$. 

Using this results, we see that for an initial state $\ket{\mu}$, the oracle unitary acts in the same way as the unitary $U'_{\boldsymbol a} = \sum_{\boldsymbol{x}}(-1)^{\boldsymbol{a}\cdot\boldsymbol{x}}\ket{\psi'_{\boldsymbol{x}}} \! \bra{\psi'_{\boldsymbol{x}}}$, which is diagonal in the basis $\{\ket{\psi'_{\boldsymbol{x}}}\}_{\boldsymbol{x}}$. Thus, we can estimate the optimal performance by evaluating the maximal probability to distinguish the unitaries $U'_{\boldsymbol a}$, when applied on the state $\ket{\mu}$. In full analogy to Eq.~(\ref{performance-d}) we obtain 
\begin{equation}
P (\ket{\mu})= \frac{1+R'(\ket{\mu}\!\bra{\mu})}{2^{N}}, \label{eq:PerformanceNewBasis}
\end{equation}
where $R'$ is the robustness of coherence in the basis $\{\ket{\psi'_{\boldsymbol{x}}}\}_{\boldsymbol{x}}$. 

Using the structure of the states $\ket{\psi'_{\boldsymbol{x}}}$ and the properties of the robustness of coherence, this result can be expressed in terms of the robustness of coherence $R$ with respect to the computational basis:
\begin{eqnarray}
   P (\ket{\mu}) &&= \frac{1}{2^{N}} \Bigg[1+|b|^2R(\ket{\phi}\!\bra{\phi}) \nonumber\\
   &&+2 |b|\sum_{x\neq \boldsymbol{0}}|c_{\boldsymbol{x}}|\Bigg(\sqrt{1-|b|^2\bigg(1-|c_{\boldsymbol{0}}^2|\bigg)}-|b||c_{\boldsymbol{0}}|\Bigg)\Bigg].
\end{eqnarray}
This completes the proof of Theorem~\ref{theorem_1}.

As we see from the Eq.~(\ref{eq:PerformanceNewBasis}), if we want to reach to the maximum performance $P=1$, we must have $R(\ket{\mu}\!\bra{\mu})=2^N-1$. Thus, $\ket{\mu}$ must be maximally coherent state in the basis $\{\ket{\psi'_{\boldsymbol{x}}}\}_{\boldsymbol{x}}$, which implies
\begin{align}
    \textrm{   }|b|^2|c_{\boldsymbol{x}}|^2=\frac{1}{2^N} \,\,\,\forall \boldsymbol{x}\neq \boldsymbol{0}, \label{AllCoeff}\\
    |a|^2+|b|^2 |c_{\boldsymbol{0}}|^2=\frac{1}{2^N}.
\end{align}
As $\sum_{\boldsymbol{x}\neq\boldsymbol{0}}c_{\boldsymbol{x}}=1-|c_{\boldsymbol{0}}|^2$ and $|a|^2=1-|b|^2$, we can rewrite the above equations as follow:
\begin{align}
    |b|^2|c_{\boldsymbol{0}}|^2=\frac{1}{2^N}\\
    1-|b|^2(1-|c_{\boldsymbol{0}}|^2)=\frac{1}{2^N}.
\end{align}
Solving these two equations for $|b|$ and $|c_{\boldsymbol{0}}|$, we obtain $|b|=1$ and $c_{\boldsymbol{0}}=\frac{1}{\sqrt{2^N}}$. From Eq. (\ref{AllCoeff}) we also have $|c_{\boldsymbol{x}}|=\frac{1}{\sqrt{2^N}}$ for all $\boldsymbol{x}$. This implies that in order to have the maximum performance, the initial state must be $\ket{\mu}=\ket{-}\ket{\psi_{\max}}$ where $\ket{\psi_{\max }}$ is a maximally coherent state in the computational basis.

Furthermore, we can generalize the result of Theorem~\ref{theorem_1} for mixed states. If we initiate the BV algorithm in a state of the form $\rho=\sum_{i}p_{i}\ket{\mu_{i}}\!\bra{\mu_{i}}$ with $\ket{\mu_{i}}=a_{i}\ket{+}\ket{\phi}+b_{i}\ket{-}\ket{\psi_{i}}$, $|a_{i}|^2+|b_{i}|^2=1$, and $\langle00...0|\psi_{i}\rangle=0$, the action of the oracle unitary on this state is
\begin{equation}
    U=\sum_{\boldsymbol{x}\neq\boldsymbol{0}}(-1)^{\boldsymbol{a}\cdot\boldsymbol{x}}\ket{-}\!\bra{-}\otimes\ket{\boldsymbol{x}}\!\bra{\boldsymbol{x}}+\ket{+}\!\bra{+}\otimes\ket{\phi}\!\bra{\phi}.
\end{equation}
In this case and in the basis $\{\ket{+}\ket{\phi}\}\cup \{\ket{+}\ket{\boldsymbol{x}}\}_{\boldsymbol{x}\neq \boldsymbol{0}}$, the action of $U$ is the same as the action of the oracle unitary on the system qubit in the computational basis when the oracle register is in the state $\ket{-}$. Hence, by Eq.~(\ref{performance-d}) we have:
\begin{equation}
    P(\rho)=\frac{1+R'(\rho)}{2^N}
\end{equation}
in which $R'(\rho)$ is the robustness of coherence in the basis $\{\ket{+}\ket{\phi}\}\cup \{\ket{+}\ket{\boldsymbol{x}}\}_{\boldsymbol{x}}$.

\subsection{Classical probabilistic BV algorithm}
In this section we prove the result in Eq.~(\ref{Classical-Performance}) of the main text. For the case  $\boldsymbol{x}=\boldsymbol{0}$ we have $f(\boldsymbol x) = 0$, which means that the oracle does not imprint any information onto the bit string $(i,\boldsymbol x)$. Thus, the probability of success in this case is equal to $1/2^N$. Now we prove that for any $r\in\{0,1\}$ and any string $\boldsymbol{x}\neq \boldsymbol{0}$ with $x_{i}\in \{0,1\}$, there are $2^{N-1}$ number of strings $\boldsymbol{a}$ with $a_{i}\in \{0,1\}$ such that $\boldsymbol{a}\cdot\boldsymbol{x} \textrm{ mod }2=r$. Note that $N$ is the length of the strings $\boldsymbol{a}$ and $\boldsymbol{x}$. We have,
\begin{equation}
   \boldsymbol{a}\cdot\boldsymbol{x} \textrm{ mod }2= \left(\sum_{i}a_{i}x_{i}\textrm{ mod }2\right)\textrm{ mod }2.
\end{equation}
As $\boldsymbol{x}\neq \boldsymbol{0}$, there is at least one $x_{l}= 1$ ($l\in\{1,2,..,N\}$). Therefore,
\begin{equation}
    \boldsymbol{a}\cdot\boldsymbol{x} \textrm{ mod }2 = \left(\left[\sum_{i\neq l} a_{i}x_{i}\right]\textrm{ mod }2 + a_l\right) \textrm{ mod }2=r.
\end{equation}
From this result we see that there are $2^{N-1}$ different bit strings $\boldsymbol{a}$ for a given $r$. Thus, knowing $(i \oplus f(\boldsymbol{x}),\boldsymbol{x})$ for any $\boldsymbol{x}\neq \boldsymbol{0}$, the success probability for correctly guessing $\boldsymbol a$ is equal to $1/2^{N-1}$. 

\subsection{Avoiding entanglement in BV algorithm}

We prove that for the probabilistic BV algorithm to exhibit performance above $1/2^N$ without entanglement in the initial and the final state, the total initial state must have the form
\begin{equation}
\ket{\mu}=\ket{-}\ket{\phi} \label{eq:NoEntanglement2}
\end{equation}
with an $N$-qubit product state $\ket{\phi}$. recall that any pure initial state of $N+1$ qubits can be written as $\ket{\mu} = a \ket{+}\ket{\phi'} + b \ket{-}\ket{\phi}$. After the action of the oracle unitary $U_{\boldsymbol a}$ the state takes the form 
\begin{equation}
U_{\boldsymbol a}\ket{\mu}= a \ket{+}\ket{\phi'} + b \ket{-}\ket{\psi_{\boldsymbol a}} \end{equation}
with $\ket{\psi_{\boldsymbol a}} = \sum_{\boldsymbol{x}}c_{\boldsymbol{x}}(-1)^{\boldsymbol{a}\cdot\boldsymbol{x}}\ket{\boldsymbol{x}}$. Tracing out the oracle register, the reduced state of the $N$-qubit system takes the form
\begin{equation}
\rho_{\boldsymbol a}=|a|^{2} \ket{\phi'}\!\bra{\phi'}+|b|^{2}\ket{\psi_{\boldsymbol a}}\!\bra{\psi_{\boldsymbol a}}.
\end{equation}
For the state $U_{\boldsymbol a} \ket{\mu}$ to be separable for all bit strings $\boldsymbol a$, it must be either $|a|=1$ or $|b|=1$, or $\ket{\psi_{\boldsymbol a}} = \ket{\phi'}$ for all $\boldsymbol a$. In the latter case the final state $U_{\boldsymbol a} \ket{\mu}$ does not depend on $\boldsymbol a$, which means that the performance will be minimal in this case. The same is true if $|a|=1$. The only remaining case is $|b| = 1$, which proves that the initial state must be of the form~(\ref{eq:NoEntanglement2}).

The arguments presented above show that the only possibility to avoid entanglement in the initial and the final state of the algorithm (while keeping nontrivial performance) is to initialize the algorithm in a state of the form $\ket{-}\ket{\phi}$, with a product state $\ket{\phi}$.

\subsection{Optimal POVM in the absence of entanglement}
\label{app:opt_POVM}

Here, we prove that in in the case of qubits initialized in $\ket{\mu}=\ket{-}\ket{\phi}$, where $\ket{\phi}=\otimes_{i=1}^N \ket{\phi^i}$, it is possible to achieve the maximum in Eq.~\eqref{eq:Psucc_oracle} with non-entangling measurements. For this, we consider a POVM with elements $M_{\boldsymbol{a}} = \bigotimes_{i=1}^N M^{(i)}_{a_i}$, where $\{M^{(i)}_{a_i}\}$ is a single-qubit POVM acting on $i$-th qubit.

Using the fact that the action of the oracle on $\ket{-}\ket{\phi}$ is to implement the unitary $V_{\boldsymbol a} = \otimes_{i=1}^N \sigma_{z,i}^{a_i}$ on $\ket{\phi}$, we have:

\begin{align}
    \frac{1}{2^N} \sum_{\boldsymbol{a}} \tr\left[U_{\boldsymbol{a}} \ket{\mu}\!\bra{\mu} U_{\boldsymbol{a}}^{\dagger} M_{\boldsymbol{a}}\right] &=\prod_{i=1}^N \frac{1}{2} \sum_{a_i} \tr\left[ \sigma_{z,i}^{a_i} \ket{\phi^i}\!\bra{\phi^i} \sigma_{z,i}^{a_i} M^{(i)}_{a_i}\right].
\end{align}
Now, as $P(\ket{\mu}\!\bra{\mu})$ is the maximum over all POVMs of the right-hand side of Eq.~\eqref{eq:Psucc_oracle}, we have:
\begin{align}
    P(\ket{\mu}\!\bra{\mu}) \geq \max_{\{M^{(i)}_{a_i}\}} \prod_{i=1}^N \frac{1}{2} \sum_{a_i} \tr\left[\sigma_{z,i}^{a_i} \ket{\phi^i}\!\bra{\phi^i} \sigma_{z,i}^{a_i} M^{(i)}_{a_i}\right].
    \label{eq:app:Prho_ineq}
\end{align}
Our goal is now to show that the inequality~\eqref{eq:app:Prho_ineq} is actually an equality. In order to show it, we recall that $P(\ket{\mu}\!\bra{\mu})=[1+R(\ket{\phi}\!\bra{\phi})]/2^N$, see Eq.~\eqref{eq:BVQubits} of the main text. Using the fact that for pure states the robustness of coherence and the $\ell_1$-norm of coherence coincide~\cite{PCBNJA_16}, and using the properties of $\ell_1$-norm of coherence~\cite{BU20171670} the following equality is true:
\begin{align}
    1+R(\ket{\phi}\!\bra{\phi})=\prod_{i=1}^N [1+R(\ket{\phi^i}\!\bra{\phi^i})].
    \label{eq:app:R}
\end{align}

Now, we can choose ${M^{(i)}_{a_i}}$ which satisfies for any $i \in [1,N]$: 
\begin{align}
    \frac{1}{2} \max_{\{M^{(i)}_{a_i}\}} \sum_{a_i} \tr[\sigma_z^{a_i} \ket{\phi^i}\!\bra{\phi^i} \sigma_z^{a_i} M^{(i)}_{a_i}]&=P(\ket{-}\ket{\phi^i}) 
    \\&=\frac{1+R(\ket{\phi^i}\!\bra{\phi^i})}{2}. \nonumber
\end{align}
This results comes from the fact that the BV algorithm in the pure single qubit case has its maximal performance equal to $[1+R(\ket{\phi^i}\!\bra{\phi^i})]/2$.
Finally, as:
\begin{align}
    P(\ket{\mu}\!\bra{\mu}) &\geq \max_{\{M^{(i)}_{a_i}\}} \prod_{i=1}^N \frac{1}{2} \sum_{a_i} \tr\left[\sigma_{z,i}^{a_i} \ket{\phi^i}\!\bra{\phi^i} \sigma_{z,i}^{a_i} M^{(i)}_{a_i}\right]\notag \\
    & \geq \prod_{i=1}^N  \frac{1}{2} \max_{\{M^{(i)}_{a_i}\}} \sum_{a_i} \tr\left[ \sigma_{z,i}^{a_i} \ket{\phi^i}\!\bra{\phi^i} \sigma_{z,i}^{a_i} M^{(i)}_{a_i}\right],
\end{align}
we deduce:
\begin{align}
    P(\ket{\mu}\!\bra{\mu}) \geq \prod_{i=1}^N \frac{1+R(\ket{\phi^i}\!\bra{\phi^i})}{2} = \frac{1+R(\ket{\phi}\!\bra{\phi})}{2^N}
\end{align}
As we know that $P(\ket{\mu}\!\bra{\mu})=(1+R(\ket{\phi}\!\bra{\phi}))/2^N$, we provided a concrete example of a non-entangling POVM that allows to maximize Eq.~\eqref{eq:Psucc_oracle} for the case that the system is in a pure product states of $N$ qubits.
\\

\subsection{W-states cannot be maximally coherent}

Here we prove that W-states with the number of particles $N\geq 3$ can never be maximally coherent. Let us consider the most general form of W-state in a given $N$-qubit system \cite{dur2000three}: 
\begin{align}
    \ket{\Psi_W} = \frac{1}{\sqrt{N}}(e^{i\phi_{1}}\ket{\psi_1}\ket{\psi_2}..\ket{\psi_N^\perp}+e^{i\phi_{2}}\ket{\psi_1}..\ket{\psi_{N-1}^\perp}\ket{\psi_N}\nonumber\\
    +...+e^{i\phi_{N}}\ket{\psi_1^\perp}\ket{\psi_2}..\ket{\psi_N})\nonumber\\
    =\frac{1}{\sqrt{N}} [\ket{\psi_1}(e^{i\phi_{1}}\ket{\psi_2}..\ket{\psi_N^\perp}+..+e^{i\phi_{N-1}}\ket{\psi_2^\perp}..\ket{\psi_N})\nonumber\\
    +\ket{\psi_1^\perp}e^{i\phi_{N}}\ket{\psi_2}..\ket{\psi_N}]
\end{align}
with $\{\ket{\psi_i}, \ket{\psi_i^\perp}\}_{i=1}^N$ forming a basis and $\phi_{i}$ are some phases. Let us now define
\begin{align}
\ket{\Phi}_{N-1} & =\frac{1}{\sqrt{N-1}}\left(e^{i\phi_{1}}\ket{\psi_{2}}\ldots\ket{\psi_{N}^{\perp}}+\ldots+e^{i\phi_{N-1}}\ket{\psi_{2}^{\perp}}\ldots\ket{\psi_{N}}\right), \nonumber \\
\ket{\Phi^{\perp}}_{N-1} & =e^{i\phi_{N}}\ket{\psi_{2}}..\ket{\psi_{N}}.
\end{align}
Note that $\ket{\Phi^\perp}_{N-1}$ is orthogonal to the state $\ket{\Phi}_{N-1}$. Hence $\Ket{\Psi_W}$ can be expressed as
\begin{equation}
    \ket{\Psi_W}= \frac{\sqrt{N-1}}{\sqrt{N}} \ket{\psi_1}\ket{\Phi}_{N-1}+ \frac{1}{\sqrt{N}}\ket{\psi_1^\perp}\ket{\Phi^\perp}_{N-1}.
\end{equation}
Substituting 
\begin{align}
\ket{\psi_{1}} & =a\ket{0}+b\ket{1},\\
\ket{\psi_{1}^{\perp}} & =b^{*}\ket{0}-a^{*}\ket{1}
\end{align}
with $|a|^2+|b|^2=1$ we further obtain
\begin{align}
    \ket{\Psi_W}= \ket{0}\left(\frac{\sqrt{1-N}}{\sqrt{N}} a \ket{\Phi}_{N-1}+\frac{1}{\sqrt{N}} b^* \ket{\Phi^\perp}_{N-1}\right) \label{eq:WstateProof-1}\\
    +\ket{1}\left(\frac{\sqrt{1-N}}{\sqrt{N}} b \ket{\Phi}_{N-1}-\frac{1}{\sqrt{N}} a^* \ket{\Phi^\perp}_{N-1}\right). \nonumber
\end{align}

Note that any $N$-qubit maximally coherent state $\ket{\psi_{\max,N}}$ can be written as 
\begin{equation}
\ket{\psi_{\max,N}}=\frac{1}{\sqrt{2}}\left(\ket{0}\ket{\psi_{\max,N-1}}+\ket{1}\ket{\psi'_{\max,N-1}}\right), \label{eq:WstateProof-2}
\end{equation}
where $\ket{\psi_{\max,N-1}}$ and $\ket{\psi'_{\max,N-1}}$ are $(N-1)$-qubit maximally coherent states. Comparing Eqs.~(\ref{eq:WstateProof-1}) and (\ref{eq:WstateProof-2}), we see that for a W-state to be maximally coherent, it must be that the (unnormalized) states $\frac{\sqrt{1-N}}{\sqrt{N}} a \ket{\Phi}_{N-1}+\frac{1}{\sqrt{N}} b^* \ket{\Phi^\perp}_{N-1}$ and $\frac{\sqrt{1-N}}{\sqrt{N}} b \ket{\Phi}_{N-1}-\frac{1}{\sqrt{N}} a^* \ket{\Phi^{\perp}}_{N-1}$ have the same norm. Evaluating the norm for these two states and equating them, we obtain
\begin{equation}
    (1-N)|a|^2+|b|^2=(1-N)|b|^2+|a|^2.
\end{equation}
This equation implies that $|a|^2=|b|^2$. We have the exact same reasoning for other qubits. Hence, for $\ket{\Psi_{W}}$ to be a maximally coherent state, it is necessary to have 
\begin{align}
\ket{\psi_{i}} & =\frac{\ket{0}+e^{i\theta_{i}}\ket{1}}{\sqrt{2}},\\
\ket{\psi_{i}^{\perp}} &   =\frac{\ket{0}-e^{i\theta_{i}}\ket{1}}{\sqrt{2}}
\end{align}
with some phases $\theta_{i}$. Now we consider the following $N$-qubit W-state:
\begin{align}
    \ket{W}=\frac{1}{\sqrt{N}}\sum_{j=1}^{N}e^{i\phi_{j}}\ket{(+)^{N-1},(-)_{j}}
\end{align}
with $\ket{(+)^{N-1},(-)_{j}}=\ket{+}_1\ket{+}_2...\ket{-}_j...\ket{+}_N$ and $\ket{\pm}_j=\frac{\ket{0}\pm e^{i\theta_{j}}\ket{1}}{\sqrt{2}}$. We also define $\ket{1_{k}}=e^{i\theta_{k}}\ket{1}$ and $\ket{0_{k}}=\ket{0}$. We want to see if $\ket{W}$ is a maximally coherent state in the computational basis. If it is a maximally coherent state then all the states in the computational basis occur with the same probability. Since $\ket{(0)^{N-1},1_{j}}=\ket{0}_{1}\ket{0}_{2}...\ket{1}_{j}...\ket{0}_{N}$ are the same vectors in the computational basis but with different phases, provided that $\ket{W}$ is a maximally coherent state, they must have the coefficients with the magnitude of $\frac{1}{\sqrt{2^N}}$ while the state $\ket{W}$ is expanded in the computational basis. Expanding $\ket{W}$ in the computational basis, we denote the coefficient of the state $\ket{\boldsymbol{x}}$ with $f(\ket{\boldsymbol{x}})$ i.e. $\ket{W}=\sum_{\boldsymbol{x}}f(\ket{\boldsymbol{x}})\ket{x}$. Let us first evaluate $f(\ket{(0)^{N-1},1_{k}})$ and $f(\ket{0}^{\otimes N})$:
\begin{align}
    f(\ket{0}^{\otimes N})&=\frac{1}{\sqrt{N2^N}}\sum_{j=1}^{N}e^{i\phi_{j}},\\
    f(\ket{(0)^{N-1},1_{k}})&=\frac{1}{\sqrt{N2^N}}(\sum_{j=1}^{N}e^{i\phi_{j}}-2e^{i\phi_{k}}).
\end{align}
If $\ket{W}$ is a maximally coherent state, we must have:
\begin{align}
    f(\ket{0}^{\otimes N})&=\frac{1}{\sqrt{N2^N}}\sum_{j=1}^{N}e^{i\phi_{j}}=\frac{e^{i\alpha_{0}}}{\sqrt{2^N}}, \\
    f(\ket{(0)^{N-1},1_{k}})&=\frac{1}{\sqrt{N2^N}}\left(\sum_{j=1}^{N}e^{i\phi_{j}}-2e^{i\phi_{k}}\right)=\frac{e^{i\alpha_{k}}}{\sqrt{2^N}}.
\end{align}
For some $\alpha_{k}$ and $\alpha_{0}$. As only the relative phases are important, we set the phase $\alpha_{0}=0$ and we get the following equations:
\begin{align}
   \sum_{j=1}^{N}e^{i\phi_{j}}&=\sqrt{N},\\
   \sum_{j=1}^{N}e^{i\phi_{j}}-2e^{i\phi_{k}}&=\sqrt{N}e^{i\alpha_{k}}.
\end{align}
We can solve these set of equations for all $\phi_{k}$. Substituting the first equation in the second one and simplifying it, we obtain:
\begin{align}
    \sqrt{N}\frac{1-e^{i\alpha_{k}}}{2}=e^{i\phi_{k}}\notag\\
    \Longleftrightarrow \sqrt{\frac{N}{2}}\sqrt{1-\cos{\alpha_{k}}}e^{i\arctan\left({\frac{-\sin{\alpha_{k}}}{1-\cos{\alpha_{k}}}}\right)}=e^{i\phi_{k}}
\end{align}
These equations imply that $\cos{\alpha_{k}}=1-\frac{2}{N}$ and
\begin{equation}
    \phi_{k}=\pm \arctan{\frac{|\sin{\alpha_{k}}|}{|1-\cos{\alpha_{k}}|}}= \pm \arctan\sqrt{N-1}.
\end{equation} 
Now, let us calculate $f(\ket{0}^{\otimes N-2}\otimes \ket{1}\otimes \ket{1})$:
\begin{align}
    f(\ket{0}^{\otimes N-2}\otimes \ket{1}\otimes \ket{1})=\frac{1}{\sqrt{N2^N}}\left(\sum_{j=1}^{N}e^{i\phi_{j}}-2e^{i\phi_{N-1}}-2e^{i\phi_{N}}\right).\label{Coeff011}
\end{align}
If the $\ket{W}$ state is a maximally coherent state, this coefficient should also be equal to $\frac{e^{i\theta}}{\sqrt{2^N}}$ for some phase $\theta$. Substituting $\phi_{k}=\pm \arctan\sqrt{N-1}$ and $ \sum_{j=1}^{N}e^{i\phi_{j}}=\sqrt{N}$ in Eq.~\eqref{Coeff011} and equating the coefficient with $\frac{e^{i\theta}}{\sqrt{2^N}}$, we have:
\begin{align}
    \frac{1}{\sqrt{2^N}}\left[1-\frac{2}{\sqrt{N}}\left(e^{\pm i\arctan\sqrt{N-1}}+e^{\pm i\arctan\sqrt{N-1}}\right)\right]=\frac{e^{i\theta}}{\sqrt{2^N}}.
\end{align}
Simplifying the above equation, we obtain:
\begin{align}
    1-\frac{2}{\sqrt{N}}\left(e^{\pm i\arctan\sqrt{N-1}}+e^{\pm i\arctan\sqrt{N-1}}\right)=e^{i\theta}.
\end{align}
The last equation does not have any solutions for $N\in \mathbb{N}$ and $N>2$. This proves that the magnitude of coefficients of the states $\ket{0}^{\otimes N-2}\otimes \ket{1,1}$ and $\ket{(0)^{N-1},1_{k}}$ when we expand the $\ket{W}$ state in the computational basis, cannot be the same and equal to $\frac{1}{\sqrt{2^N}}$. This proves that a W-state cannot be maximally coherent for $N>2$.

\subsection{Proof of Theorem \ref{thm:Pseudopure}}

In order to prove the theorem, first we show that for any pseudo-pure maximally coherent state, the the robustness of coherence coincides with the $\ell_{1}$-norm of coherence, defined as $C_{\ell_1}(\rho) = \sum_{i \neq j}|\rho_{ij}|$~\cite{BaumgratzPhysRevLett.113.140401}. Consider that the pseudo-pure maximally coherent state, which we denote here by $\rho_{s}$, is defined as 
\begin{equation}
\rho_{s}=p \ket{\psi_{\max}}\bra{\psi_{\max}}+(1-p)\frac{\openone}{d},
\end{equation}
in which $d$ is the dimension of the Hilbert space, $ 0\leq p \leq 1$, and $\ket{\psi_{\max}}$ is a maximally coherent state. For $C_{\ell_{1}}(\rho_{s})$we have:
\begin{equation}
    C_{\ell_{1}}(\rho_{s})=\sum_{x,y,x\neq y}|\rho_{s,xy}|= p (d-1).
\end{equation}
Here, $\rho_{s,xy}$ are the elements of the density matrix of $\rho_s$ in the basis $\{\ket{x}\}$. We use the SDP form of the robustness of coherence to evaluate $R(\rho_{s})$:
\begin{align}
R(\rho_{s}) & =\max_{X}\tr(\rho_{s}X)-1\\
 & =\left[p \braket{\psi_{\max}|X^{*}|\psi_{\max}}+(1-p)\frac{1}{d}\tr(X^{*})\right]-1,\nonumber 
\end{align}
where $X^{*}$ is the matrix maximizing the SDP. 

Referring to Eq.~\eqref{XDiagElement}, we know that the diagonal elements of the matrix $X^{*}$ are the same and are equal to $1$ and also it is a positive matrix, hence we can write $X^{*}\equiv d\rho_{x}$ with some quantum state $\rho_x$. We further obtain:
\begin{align}
\max_{X}\tr(\rho_{s}X)-1 & =d p \braket{\psi_{\max}|\rho_{x}|\psi_{\max}}-p.
\label{BoundedPurity-SDP1}
\end{align}
Note that the restriction on $X^*$ only demands that all the diagonal elements of $\rho_x$ are the same. Now, the term $\bra{\psi_{\max}}\rho_{x}\ket{\psi_{\max}}$ in Eq. (\ref{BoundedPurity-SDP1}) is maximum when we consider $\rho_{x}$ as the maximally coherent state $\ket{\psi_{\max}}\!\bra{\psi_{\max}}$, which fulfils the criterion of $X^*$. So it can be seen as, $X^*=d\ket{\psi_{\max}}\!\bra{\psi_{\max}}$ maximizes the SDP for $\rho_{s}$. Thus
\begin{equation}
    R(\rho_{s})=p (d-1)=C_{\ell_{1}}(\rho_{s}).
    \label{RobustforPseudo}
\end{equation}

Now we use Lagrange multipliers method to maximize the value of $\ell_{1}$-norm of coherence for a given amount of purity. The purity of the density matrix $\rho$ can be expressed in terms of the absolute value of the components $|\rho_{i,j}|$ as below:
\begin{equation}
    \tr(\rho^2)=\sum_{i,j}|\rho_{i,j}|^2.
\end{equation}
The goal is to maximize the $\ell_1$-norm of coherence with $\gamma$ amount of purity. Note that, instead of maximizing $C_{\ell_{1}}$, we maximize the function $g=\sum_{i,j}|\rho_{i,j}|=C_{\ell_{1}}+1$ with the constraint $\sum_{i=j}\rho_{i,j}=1$ which implies maximizing $C_{\ell_{1}}$ function.

Therefore, our maximization problem is as follows:
\begin{itemize}
    \item Constraint 1: $C_{1}=\sum_{i,j}|\rho_{i,j}|^2-\gamma=0$.
    
    \item Constraint 2: $C_{2}=\sum_{i=j}\rho_{i,j}-1=0$. 
    \item $\lambda_{1}$ and $\lambda_{2}$ are the Lagrange multipliers corresponding to $C_{1}$ and $C_{2}$ constraints respectively.
    
    \item The function $g=\sum_{i,j}|\rho_{i,j}|$ which is aimed to be maximized with respect to the variables $|\rho_{i,j}|$.
\end{itemize}
Note that we also have two other constraints which are the hermiticity and positivity of the density matrix $\rho$. We do not apply these constraints during the maximization, but we will check them in the end for the maximizing state.

Applying Lagrange multipliers method, we obtain the following equations:
\begin{align}
    \frac{dg}{d|\rho_{i,j}|}-\lambda_{1}\frac{dC_{1}}{d|\rho_{i,j}|}-\lambda_{2}\frac{dC_{2}}{d|\rho_{i,j}|}=0.
\end{align}
Simplifying these equations and considering the constraints, we have the following set of equations:
\begin{align}
    \mathrm{for}~i\neq j, 1-2\lambda_{1}|\rho_{i,j}|=0, \\
    \mathrm{for}~i=j=k, 1-2\lambda_{1}|\rho_{k,k}|+\lambda_{2}=0,\\
    C_{1}=\sum_{i,j}|\rho_{i,j}|^2-\gamma=0,\\
    C_{2}=\sum_{i=j}|\rho_{i,j}|-1=0.
\end{align}
Solving these equations for $|\rho_{i,j}|$, $\lambda_{1}$ and $\lambda_{2}$ results in
\begin{align}
    i\neq j, |\rho_{i,j}|=\frac{1}{2\lambda_{1}},\\
    i=j=k, |\rho_{k,k}|=\frac{1+\lambda_{2}}{2\lambda_{1}}.
\end{align}
Considering a $d$-dimensional system, we obtain,
\begin{align}
    \lambda_{2} &= \frac{\sqrt{1-\frac{1}{d}}}{\sqrt{\gamma-\frac{1}{d}}}-1,\\
    \lambda_{1} &= \frac{d\sqrt{1-\frac{1}{d}}}{2\sqrt{\gamma-\frac{1}{d}}}.
\end{align}
As $|\rho_{i,j}|=|\rho_{j,i}|$ and also we have the freedom to choose the phases in $\rho_{i,j}=|\rho_{i,j}|e^{i\phi_{i,j}}$, we can choose $\phi_{i,j}$ so that $\rho_{\max,\gamma}$ is Hermitian.

Using $\rho_{i,j}$ obtained, we can write the maximizing state $\rho_{\max,\gamma}$ in the following form:
\begin{equation}
    \rho_{\max,\gamma}=\frac{d}{2\lambda_{1}}\ket{\psi_{\max}}\!\bra{\psi_{\max}}+\frac{\lambda_{2}}{2\lambda_{1}}\openone.
    \label{RhoMax}
\end{equation}
As $\lambda_{1},\lambda_{2}\geq 0$ and $\frac{d}{2\lambda_{1}}+\frac{d\lambda_{2}}{2\lambda_{1}}=1$, $\rho_{\max,\gamma}$ is a valid density matrix and maximizes the $\ell_{1}$-norm of coherence with bounded purity of $\gamma$. Also, the maximum amount of coherence is 
\begin{equation}
    C_{\ell_{1},\max}=\frac{d^2-d}{2\lambda_{1}}.
\end{equation}
As we already proved that the amount of $\ell_{1}$-norm of coherence coincides with the amount of robustness of coherence for any pseudo-pure maximally coherent state and $R(\rho)\leq C_{\ell_{1}}(\rho)$ for any $\rho$ \cite{PCBNJA_16,NBCPJA_16}, the state $\rho_{\max,\gamma}$ in Eq. (\ref{RhoMax}) also maximizes $R(\rho)$ with the same amount as $C_{\ell_{1},\max}$, given the purity $\tr(\rho^2)=\gamma$:
\begin{align}
    R_{\max}=\frac{d(d-1)}{2\lambda_{1}}.
\end{align}

\subsection{Probabilistic BV algorithm for qudits}

In the BV algorithm with qudits, the goal is to learn the string $\boldsymbol k$ with $k_{i}\in \{1,...,D\}$ encoded into the linear function
\begin{equation}
    f(\boldsymbol{x})=\boldsymbol{k} \cdot \boldsymbol{x} \textrm{ mod }D=\sum_{i=1}^{N}k_{i}x_{i}\textrm{ mod }D,
\end{equation}
where $x_{i}\in \{1,...,D\}$. Similar to the qubit version of the algorithm, we assume that the function is encoded into an oracle unitary acting as
\begin{equation}
    U_{\boldsymbol k}\ket{j}\ket{\boldsymbol{x}}=\ket{j+f(\boldsymbol{x})\textrm{ mod }D }\ket{\boldsymbol{x}}
\end{equation}
with $j\in \{0,...,D-1\}$. As we will now see, there will be no entanglement between the oracle register and the qudit system  if the oracle unitary is applied onto a state of the form $\ket{-_D}\ket{\phi}$ with
\begin{equation}
    \ket{-_D}=\frac{1}{\sqrt{D}}\sum_{k=0}^{D-1}e^{-i\frac{2\pi}{D}k}\ket{k},
\end{equation}
and $\ket{\phi}$ is a product state of $N$ qudits. Moreover, the performance of the protocol, when applied onto states of the form $\ket{-_D}\!\bra{-_D}\otimes \rho$ can be evaluated as follows: 
\begin{equation}
    P(\ket{-_D}\bra{-_D} \otimes \rho)=\frac{1+R(\rho)}{d},
    \label{BV-QDit}
\end{equation}
where $R(\rho)$ is the robustness of coherence in the computational basis and $d=D^N$. 

Having the relation in Eq. (\ref{BV-QDit}), similar to the Theorem \ref{thm:Pseudopure} we can also show that if the state of the oracle register is $\ket{-_D}$ then the state of system qudits maximizing the performance $P(\rho)$ with bounded purity $\textrm{Tr}(\rho)=\gamma$, must be of the form
\begin{equation}
    \rho = p \ket{\psi_{\max}}\!\bra{\psi_{\max}}+(1-p)\frac{\openone}{D^N}
\end{equation}
with $0\leq p \leq 1$ and $\ket{\psi_{\max}}$ is an $N$-qudit maximally coherent state. The proofs of the aforementioned results for qudits are very similar to the proofs of the corresponding ones for qubits that have been discussed before.

First, we will prove that in the probabilistic BV algorithm with qudits, if the oracle register is in the $\ket{-_D}$ state then after the action of the oracle, there will be no entanglement between the oracle register and the system qudits and also $U_{\boldsymbol{k}}$ will act as a non-entangling gate. Let $\ket{\boldsymbol x}$ be an arbitrary vector in the computational basis. We have:
\begin{equation}
    U_{\boldsymbol{k}}\ket{-_{D}}\ket{\boldsymbol{x}}=\frac{1}{\sqrt{D}}\sum_{j=0}^{D-1}e^{-i\frac{2\pi}{D}j}U_{\boldsymbol{k}}\ket{j}\ket{\boldsymbol{x}}.
\end{equation}
The oracle unitary acts as  $U_{\boldsymbol k}\ket{j}\ket{\boldsymbol{x}}=\ket{j+f(\boldsymbol{x})\textrm{ mod }D }\ket{\boldsymbol{x}}$ with $f(\boldsymbol{x})=\boldsymbol{k}\cdot \boldsymbol{x}=\sum_{i=1}^{N}k_{i}x_{i}$. Thus, we can write:
\begin{align}
     U_{\boldsymbol{k}}\ket{-_{D}}\ket{\boldsymbol{x}}&=\frac{1}{\sqrt{D}}\sum_{j=0}^{D-1}e^{-i\frac{2\pi}{D}j}\ket{j+f(\boldsymbol{x}) \textrm{ mod }D}\ket{\boldsymbol{x}} \notag \\
     &=e^{i\frac{2\pi}{D}f(\boldsymbol{x})}\frac{1}{\sqrt{D}}\sum_{j=0}^{D-1}e^{-i\frac{2\pi}{D}j}U_{\boldsymbol{k}}\ket{j}\ket{\boldsymbol{x}} \nonumber \\
     &=e^{i\frac{2\pi}{D}f(\boldsymbol{x})}\ket{-_{D}}\ket{\boldsymbol{x}}. 
\end{align}
Thus, as we see if the state of the oracle register is $\ket{-_{D}}$, the oracle unitary acts as the non-entangling unitary $\openone\otimes\sum_{j=0}^{D-1}e^{i\frac{2\pi}{D}k_{1}j}\ket{j}\!\bra{j}\otimes \sum_{j=0}^{D-1}e^{i\frac{2\pi}{D}k_{2}j}\ket{j}\!\bra{j}\otimes \ldots \otimes \sum_{j=0}^{D-1}e^{i\frac{2\pi}{D}k_{N}j}\ket{j}\!\bra{j}$.

Now we prove the result in the Eq.~(\ref{BV-QDit}). In full analogy to the qubit case, the robustness of coherence can be evaluated with a semidefinite program, see Eqs. (\ref{eq:SDP}). In contrast to the qubit setting, we have $E(X)=\frac{1}{d}\sum_{\boldsymbol{k}}u_{\boldsymbol{k}}Xu_{\boldsymbol {k}}^{\dagger}$ and $u_{\boldsymbol{k}}=\sum_{x}e^{i\frac{2\pi}{d}\boldsymbol{k}\boldsymbol{x}}\ket{\boldsymbol{x}}\!\bra{\boldsymbol{x}}$.

In order to prove the expression in Eq.~(\ref{BV-QDit}), we will show that $P(\ket{-_D}\!\bra{-_D} \otimes \rho)$ is both lower and upper bounded by $[1+R(\rho)]/d$. As discussed above, the action of the oracle unitaries $U_{\boldsymbol k}$ onto states of the form $\ket{-_D}\!\bra{-_D} \otimes \rho$ can be expressed as
\begin{equation}
U_{\boldsymbol{k}}\ket{-_D}\!\bra{-_D}\otimes\rho U_{\boldsymbol{k}}=\ket{-_D}\!\bra{-_D}\otimes V_{\boldsymbol{k}}\rho V_{\boldsymbol{k}}^{\dagger}
\end{equation}
with the $N$-qudit unitaries
\begin{equation}
    V_{\boldsymbol{k}}=\sum_{j=0}^{D-1}e^{i\frac{2\pi}{D}k_{1}j}\ket{j}\!\bra{j}\otimes \sum_{j=0}^{D-1}e^{i\frac{2\pi}{D}k_{2}j}\ket{j}\!\bra{j}\otimes \ldots \otimes \sum_{j=0}^{D-1}e^{i\frac{2\pi}{D}k_{N}j}\ket{j}\!\bra{j}.
\end{equation}
In case of qubits this unitary will become $V_{\boldsymbol k}=\otimes_{i=1}^{N}\sigma_{z,i}^{k_{i}}$ and $k_{i}\in \{0,1\}$ \cite{krishna2016generalization}. The performance of the probabilistic BV algorithm can thus be written as
\begin{equation}
P(\ket{-_D}\!\bra{-_D} \otimes \rho)=\max_{\{M_{\boldsymbol k}\}}\frac{1}{d}\sum_{\boldsymbol k}\tr\left(V_{\boldsymbol k} \rho V_{\boldsymbol k}^{\dagger}M_{\boldsymbol k}\right).
\end{equation}
In the next step, we will show that 
\begin{equation}
\max_{\{M_{\boldsymbol k}\}}\frac{1}{d}\sum_{\boldsymbol k}\tr\left(V_{\boldsymbol k} \rho V_{\boldsymbol k}^{\dagger}M_{\boldsymbol k}\right)\geq\frac{1+R(\rho)}{d}.
\end{equation}
For this, we define the following operators:
\begin{equation}
    M_{\boldsymbol k}^{\prime}=\frac{1}{d}V_{\boldsymbol k}XV_{\boldsymbol k}^{\dagger}.
\end{equation}
Note that $X$ is the operator maximizing the SDP in Eqs.~(\ref{eq:SDP}). The operators $M_{\boldsymbol k}^{\prime}$ are positive because $X$ is positive. Since $E(X)=\openone$, we have 
\begin{equation}
\frac{1}{d}\sum_{k=1}^{d}X_{\boldsymbol{x},\boldsymbol{x}}=1
\end{equation}
where $X_{\boldsymbol{x},\boldsymbol{y}}$ are the components of the matrix $X$, thus the diagonal elements of $X$ are the same and equal to $1$. Note that as $\frac{1}{d}\sum_{\boldsymbol{k}}e^{i\frac{2\pi}{D} \boldsymbol{k}\cdot(\boldsymbol{x}-\boldsymbol{x'})}=\delta_{\boldsymbol{x},\boldsymbol{x'}}$ and $V_{\boldsymbol k}$ are diagonal unitaries, we have
\begin{align}
\sum_{\boldsymbol k}M_{\boldsymbol k}^{\prime}&=\frac{1}{d}\sum_{f}V_{\boldsymbol k}XV_{\boldsymbol k}^{\dagger} \\
&=\sum_{\boldsymbol{x},\boldsymbol{y}}\frac{1}{d}\sum_{\boldsymbol{k}}e^{i\frac{2\pi}{D}\boldsymbol{k}\cdot(\boldsymbol{x}-\boldsymbol{y})}X_{\boldsymbol{x},\boldsymbol{y}}\ket{\boldsymbol{x}}\!\bra{\boldsymbol{y}}=\openone. \nonumber
\end{align}
Therefore $\{M_{\boldsymbol k}'\}$ form a set of POVM operators. Thus we must have
\begin{align}
\frac{1}{d}\sum_{\boldsymbol{k}}\tr\left(V_{\boldsymbol{k}}\rho V_{\boldsymbol{k}}^{\dagger}M_{\boldsymbol{k}}^{\prime}\right) & =\frac{1}{d}\sum_{\boldsymbol{k}}\frac{1}{d}\tr\left(V_{\boldsymbol{k}}\rho V_{\boldsymbol{k}}^{\dagger}V_{\boldsymbol{k}}XV_{\boldsymbol{k}}^{\dagger}\right)\label{GEQ2}\\
 & =\frac{1}{d}\tr(\rho X)=\frac{1+R(\rho)}{d}\nonumber \\
 & \leq\underset{\{M_{\boldsymbol{k}}\}}{\max}~\frac{1}{d}\sum_{\boldsymbol{k}}\tr\left(V_{\boldsymbol{k}}\rho V_{\boldsymbol{k}}^{\dagger}M_{\boldsymbol{k}}\right). \nonumber 
\end{align}
Above we use the definition of the  robustness of coherence to obtain $\tr(\rho X) = 1+R(\rho)$. 

Now, we will show that the performance is also upper bounded by $\big(1+R(\rho)\big)/d$. From the definition of the robustness of coherence, it follows that 
\begin{equation}
    \rho=[1+R(\rho)]\sigma-R(\rho) \tau,
\end{equation}
with some quantum state $\tau$ and some incoherent state $\sigma$. Hence, for any POVM $\{M_{\boldsymbol a}\}$ we have,
\begin{align}
\sum_{\boldsymbol{k}}\tr\left(V_{\boldsymbol{k}}\rho V_{\boldsymbol{k}}^{\dagger}M_{\boldsymbol{k}}\right) & =[1+R(\rho)]\sum_{\boldsymbol{k}}\tr\left(V_{\boldsymbol{k}}\sigma V_{\boldsymbol{k}}^{\dagger}M_{\boldsymbol{k}}\right)\\
 & -R(\rho)\sum_{\boldsymbol{k}}\tr\left(V_{\boldsymbol{k}}\tau V_{\boldsymbol{k}}^{\dagger}M_{\boldsymbol{k}}\right)\nonumber \\
 & \leq[1+R(\rho)]\sum_{\boldsymbol{k}}\tr\left(V_{\boldsymbol{k}}\sigma V_{\boldsymbol{k}}^{\dagger}M_{\boldsymbol{k}}\right).\nonumber 
\end{align}
As $\sigma$ is an incoherent state, it holds $V_{\boldsymbol k}\sigma V_{\boldsymbol k}^{\dagger}=\sigma$ and
\begin{equation}
\sum_{\boldsymbol{k}}\tr\left(V_{\boldsymbol{k}}\sigma V_{\boldsymbol{k}}^{\dagger}M_{\boldsymbol{k}}\right)=\tr\left(\sigma\sum_{\boldsymbol{k}}M_{\boldsymbol{k}}\right)=1.
\end{equation}
Thus, we arrive at the inequality
\begin{equation}
\max_{\{M_{\boldsymbol{k}}\}}\frac{1}{d}\sum_{\boldsymbol{k}}\tr\left(V_{\boldsymbol{k}}\rho V_{\boldsymbol{k}}^{\dagger}M_{\boldsymbol{k}}\right)\leq\frac{1+R(\rho)}{d}. \label{LEQ2}
\end{equation}
From Eqs.~\eqref{GEQ2} and \eqref{LEQ2} we have
\begin{equation}
\max_{\{M_{\boldsymbol k}\}}\frac{1}{d}\sum_{\boldsymbol k}\tr\left(V_{\boldsymbol k}\rho V_{\boldsymbol k}^\dagger M_{\boldsymbol k}\right)=\frac{1+R(\rho)}{d}.
\end{equation}
This completes the proof of Eq.~(\ref{BV-QDit}).

\end{document}